\documentclass{article}

\usepackage[final]{neurips_2025_ml4ps}

\usepackage[utf8]{inputenc} 
\usepackage[T1]{fontenc}    
\usepackage{hyperref}       
\usepackage{url}            
\usepackage{booktabs}       
\usepackage{amsfonts}       
\usepackage{nicefrac}       
\usepackage{microtype}      
\usepackage{xcolor}         
\usepackage{float}
\usepackage{graphicx}
\usepackage{amsmath}
\usepackage{multirow}
\usepackage{caption}
\usepackage{subcaption}
\usepackage{wrapfig}
\usepackage{xspace}
\setcitestyle{square,numbers,sort&compress}
\newcommand{\tng}{\textsc{TNG50}\xspace}
\newcommand{\sdss}{\textsc{SDSS}\xspace}

\title{From Simulations to Surveys: Domain Adaptation for Galaxy Observations}

\author{
  Kaley Brauer \\
  Center for Astrophysics \\ Harvard University \\
  \texttt{kaley\_brauer@cfa.harvard.edu}
  \And
  Aditya Prasad Dash \\
  Physics and Astronomy\\ University of California, Los Angeles \\
  \texttt{aditya55@physics.ucla.edu}
  \And
  Meet J. Vyas \\
  International Centre for Space and Cosmology \\ Ahmedabad University \\
  \texttt{meet.v2@ahduni.edu.in}
  \And
  Ahmed Salim \\
  Department of Computing \\ Universiti Teknologi Malaysia \\
  \texttt{taha-20@live.utm.my}
  \And
  Stiven Briand Massala \\ Universit\'e Paris-Saclay, CentraleSup\'elec, ENS Paris-Saclay,\\ CNRS, LMPS - Laboratoire de M\'ecanique Paris-Saclay \\
  \texttt{massala934@gmail.com}
}

\begin{document}

\maketitle

\begin{abstract}
Large photometric surveys will image billions of galaxies, but we currently lack quick, reliable automated ways to infer their physical properties like morphology, stellar mass, and star formation rates. Simulations provide galaxy images with ground-truth physical labels, but domain shifts in PSF, noise, backgrounds, selection, and label priors degrade transfer to real surveys. We present a preliminary domain adaptation pipeline that trains on simulated \tng\ galaxies and evaluates on real \sdss\ galaxies with morphology labels (elliptical/spiral/irregular). We train three backbones (CNN, $E(2)$-steerable CNN, ResNet-18) with focal loss and effective-number class weighting, and a feature-level domain loss $\mathcal{L}_D$ built from \texttt{GeomLoss} (entropic Sinkhorn OT, energy distance, Gaussian MMD, and related metrics). We show that a combination of these losses with an OT-based “top-$k$ soft matching’’ loss that focuses $\mathcal{L}_D$ on the worst-matched source–target pairs can further enhance domain alignment. With Euclidean distance, scheduled alignment weights, and top-$k$ matching, target accuracy (macro F1) rises from $\sim$46\% ($\sim$30\%) at no adaptation to $\sim$87\% ($\sim$ 62.6\%), with a domain AUC near 0.5, indicating strong latent-space mixing. 

\end{abstract}

\begin{center}
\href{https://github.com/ahmedsalim3/galaxy-da}{https://github.com/ahmedsalim3/galaxy-da}
\end{center}

\section{Introduction}

Machine learning is now routine in the physical sciences, yet transferring models across domains remains challenging and scientifically consequential. For galaxies, simulated images carry true physical labels but differ from real observations in PSF, noise, backgrounds, selection functions, and demographics. Naively transferring such models can bias physical inferences, shifting the early/late-type mix, distorting mass–SFR demographics, or contaminating scaling relations, so robust domain adaptation must preserve calibrated, physically meaningful predictions (and uncertainties) for population studies and hypothesis testing.

Domain adaptation has emerged as a widely used and powerful paradigm for addressing classification tasks in which source and target domains follow different data distributions \cite{Shui2022,Saito2021,Wang_2018,csurka2017domainadaptationvisualapplications,Wilsongarrett2020,Sun2016CorrelationAF,Kang2019,Courty2014,FS2020,LY2022}, with successful applications across various domains \cite{Guan2021DomainAF,Tuia2016,Jakubik2023,Li2023DomainAB,Ding2025}. Astronomers have similarly adopted these techniques to infer astrophysical and cosmological parameters \cite{Vilalta_2019, Parul2024,Gilda2024,Agarwal2024,Roncoli2023,Swierc2023}. With expanding simulation capabilities and survey infrastructure, the need for robust domain-transfer methods is increasingly timely, motivating updates to techniques based on time-domain projects such as ZTF \cite{ZTFBellm2019} and Pan-STARRS1 \cite{Pan-STARRS1Flewelling2020} in preparation for projects including the Vera C. Rubin Observatory \cite{RubinDP1}, the Nancy Grace Roman Space Telescope \cite{RomanAkenson2019}, and the Euclid mission \cite{Scaramella2022}, which will provide millions of nightly alerts.

In this work, we prototype a pipeline to study how different domain-adaptation techniques can narrow this gap in a realistic setting by training on mock observations of simulated galaxies from \tng\ SKIRT \citep{RodriguezGomez_2019} and testing on real \sdss\ observations with Galaxy Zoo–derived morphology labels \citep{Lintott2008}. We frame this as a covariate- (or coverage-) shift problem in which the conditional label distribution is approximately stable, $p_S(y\mid x)\approx p_T(y\mid x)$, while the input and selection distributions differ, $p_S(x)\neq p_T(x)$, making our setup representative of many ``train on simulations, test on surveys'' applications. 
Previous studies have reported promising results \cite{Pandya2025,Pandya2023,iprijanovi2023deepastrouda,iprijanovi2022,Ciprijanovi2020}, yet room remains for methodological development. We also explore a suite of distance metrics for quantifying domain shift by updating the GeomLoss Library \cite{feydy2019interpolating} with new metrics.

\section{Data}

\begin{figure}[H]
    \centering
    \includegraphics[width=0.8\linewidth]{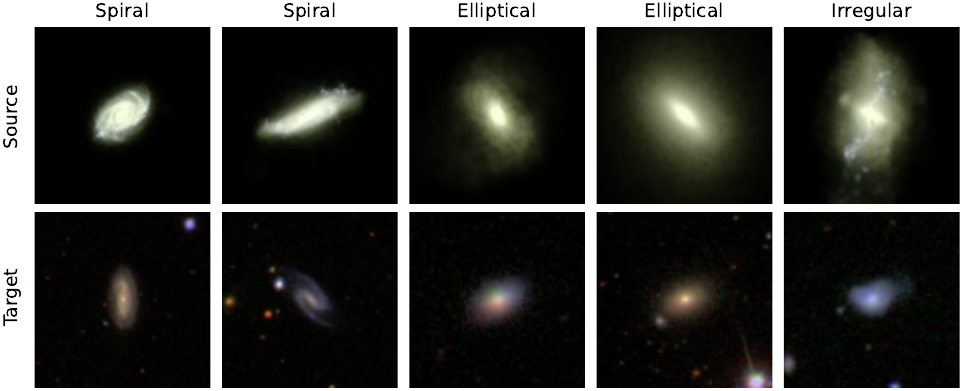}
    \caption{\textbf{Example source and target data.} Our source data are mock observations of galaxies from the \tng simulation with known galaxy properties. Our target data are real observations of galaxies. In this proof of concept, we use \sdss observations at $z \sim 0$ with known morphologies. Real data differ from simulated data in point spread function, noise, selection function, demographics, and more.}
    \label{fig:source_target}
\end{figure}

\paragraph{Source (\tng\ Simulation).} Our source set comprises mock observations of galaxies from the Illustris \tng simulation suite \citep{TNG50_1,TNG50_2}. We use SKIRT synthetic images and morphology classifications at \textit{z}=0 and \textit{z}$\approx0.05$ \citep{RodriguezGomez_2019}, restricted to \(M_\star>10^{9.5}\,M_\odot\), with quality cuts (\texttt{flag}=0, \texttt{sn\_per\_pixel}>2.5; \texttt{flag\_sersic}=0 when Sérsic parameters are used). Images have four bands (g,r,i,z) and are “idealized”; they are convolved with a Gaussian PSF and have sky noise added per \cite{RodriguezGomez_2019}. The dataset includes 3232 galaxies. We split 90/10 train/test and augment via one horizontal flip and four rotations, yielding 25{,}856 total images. For future work, we also produce labels for stellar mass, star formation rate (defining "star-forming" as specific SFR > $10^{-11}$ yr$^{-1}$), metallicity, and whether or not the galaxy includes an active galactic nuclei (AGN; galaxies with actively accreting black holes, defined as $\lambda_{\rm Edd}\simeq \dot{M}/\dot{M}_{\rm Edd}>10^{-3}$). 

\paragraph{Target (\sdss Survey).} Our target set comprises 6416 Sloan Digital Sky Survey (SDSS) galaxy images \cite{Willet2013} with morphologies labeled by Galaxy Zoo volunteers, using debiased vote fractions from \cite{Hart2016}. We select three morphology classes -- elliptical, spiral, and irregular -- by applying strict thresholds on these fractions: ellipticals require high smoothness, strong “no spiral” votes, and low edge-on probability; spirals require strong spiral-arm and feature/disk votes while excluding highly edge-on systems; irregulars require strong votes for “odd/irregular,” “merger,” or “disturbed” features. Objects flagged as artifacts are removed. Additional labels for stellar mass, star formation, and AGN come from \cite{Schawinski2010,Lintott2008,Lintott2011}; following the BPT scheme \cite{Baldwin1981} as in \cite{Schawinski2010}, we label $\mathrm{BPTCLASS}=1$ galaxies as star-forming, $\mathrm{BPTCLASS}=3$ or 4 as AGN hosts, and exclude $\mathrm{BPTCLASS}=0$ or 2.

The three morphology classes are highly imbalanced (spirals dominate, with ellipticals and especially irregulars rarer), motivating a focal loss with effective-number class weights (Section~\ref{sec:methods}). We also assume simulation-based and Galaxy~Zoo morphologies provide compatible labels for the same classes, justified by strict Galaxy~Zoo cuts and visual similarity of source and target images.

\section{Methods}
\label{sec:methods}

\paragraph{Architectures.}
We compare three backbone architectures: (i) a small convolutional network (CNN) with two conv blocks, each with two $3{\times}3$ conv-BN-ReLU layers and $2{\times}2$ max pooling, followed by a three-layer MLP classifier; (ii) an $E(2)$-steerable CNN (ESCNN) implemented with \texttt{escnn}~\cite{e2cnn} using a discrete rotation group $C_8$ and four group-equivariant convolutional blocks; and (iii) a \texttt{torchvision} \texttt{resnet18}~\cite{resnet} backbone pretrained on ImageNet~\cite{imagenet}, where we replace the final fully-connected layer by a task-specific MLP head and fine-tune only the top residual blocks, keeping the rest frozen.
Unless otherwise stated we train for 200~epochs (with early stopping) with batch size 128 using AdamW.

\paragraph{Supervised loss and class imbalance.}
All models use MLP classifier head and are trained with a supervised loss on the labeled source domain.
Because the three morphological classes are highly imbalanced, we use a focal loss $\mathcal{L}_{\text{sup}}$ in place of cross-entropy, with focusing parameter $\gamma{=}2$ and $\alpha$ determined from per-class weights computed from the “effective number of samples’’\cite{cui2019}. We also use class scales to boost minority class predictions. 
Unlike class weights (used in the loss), these scales are applied to logits to boost predictions. The scales are computed from the source training data using the formula $\text{scale}_c = \sqrt{\max(\text{count}) / \text{count}_c}$, normalized so the majority class has scale 1.0, then moderated to reduce aggressiveness. The scales are implemented as a learnable parameter in the model, initialized from these data-driven values and applied multiplicatively to the logits before the final classification output.

\paragraph{Domain adaptation objectives.}

Let $z_s$ and $z_t$ denote the feature embeddings (just before the classifier head) for source and target samples in a mini-batch.
We introduce a novel combination of domain-alignment terms $\mathcal{L}_D$ computed on L2-normalized features using differentiable distance-based losses implemented with \texttt{GeomLoss}~\cite{feydy2019interpolating}.  

In particular, we experiment with (i) entropic optimal transport (Sinkhorn divergence), (ii) the energy distance, and (iii) Gaussian MMD, corresponding to \texttt{SamplesLoss} methods ``sinkhorn'', ``energy'', and ``gaussian'' (with blur parameter $\sigma$) respectively. We extend the GeomLoss library~\cite{feydy2019interpolating} by implementing a comprehensive suite of 46 distinct distance and similarity measures derived from the taxonomy of \cite{Cha2007ComprehensiveSO}. This expansion augments the standard Euclidean formulations with metrics organized into eight distinct families: \textit{$L_p$ Minkowski, $L_1$, Intersection, Inner Product, Squared-chord, Squared $L_2$ ($\chi^2$), Shannon's Entropy}, and \textit{Combination}. 

Recognizing that the topology of the embedding space fundamentally constrains cross-domain alignment, we empirically validate the choice of metric by benchmarking model performance across 12 representative measures: \textit{inner product, cosine, Jaccard, Dice coefficient, Kumar--Hassebrook, harmonic mean, Euclidean, Manhattan, Chebyshev, Minkowski, Gower}, and \textit{average $L_1$--$L_\infty$}. These choices let us probe how different metrics on the latent space affect alignment and downstream performance.

On top of this, we add an explicit optimal-transport alignment loss that combines global OT, soft matching, and a top-$k$ penalty. 

Using $P^\lambda$, we define the full alignment loss:
\[
\mathcal{L}_{\text{OT}}(z_s,z_t)
    \;=\;
    \lambda_{\text{OT}}\, d_\lambda(p_s,p_t)
    \;+\;
    \lambda_{\text{match}}\, 
        \mathrm{MSE}\!\left(z_s,\; P^\lambda z_t\right)
    \;+\;
    \lambda_{\text{topk}}\,
    \frac{1}{k}\sum_{\ell=1}^k d_{(\ell)},
\]

where $C_{ij} = \|z_{s,i} - z_{t,j}\|_2^2$ is the pairwise cost matrix, $d_\lambda(p_s,p_t)$ is an entropically regularized OT distance computed from a Sinkhorn transport plan $P^\lambda$, $d_i=\min_j C_{ij}$ and $d_{(1)}\ge\cdots\ge d_{(k)}$ are the $k$ largest per-source closest-target distances.  

This composite formulation couples global OT alignment, soft barycentric matching, and a focused top-$k$ penalty on poorly aligned instances.
Top-$k$ losses have proved effective in supervised alignment~\cite{TopK}; here we use the same idea to emphasize the hardest-to-align source features.

Our training objective combines the supervised and domain-alignment terms as
\[
\mathcal{L}
    = \lambda_{\text{sup}}\,\mathcal{L}_{\text{sup}}
    + \lambda_D\,\mathcal{L}_D
    + \lambda_{\text{OT}}\,\mathcal{L}_{\text{OT}},
\]
where each $\lambda$ is a non-negative scalar that
controls the relative weights of the losses (with $\lambda_{\text{OT}}{=}0$ for experiments without OT/top-$k$).
We evaluate several strategies for choosing $(\lambda_{\text{sup}},\lambda_D)$:

\begin{itemize}
  \item \textbf{Baseline.} $\lambda_D{=}0$, training only with $\mathcal{L}_{\text{sup}}$.
  
  \item \textbf{Fixed-$\lambda$.} We set $\lambda_D{=}\lambda$ and $\lambda_{\text{sup}}{=}1$, tuning $\lambda$ over a small grid; $\lambda{=}0.1$ is used for the main experiments.

  \item \textbf{Trainable weights.} We use trainable parameters $\eta_1,\eta_2$ and set
  $\lambda_{\text{sup}}=\sigma(\eta_1)$, $\lambda_D=\sigma(\eta_2)$ , learning the alignment strength jointly with the network. \cite{KendallGC17}

  \item \textbf{Blur schedule + trainable weights.} For Sinkhorn-based alignment, we anneal the blur parameter from $\sigma_{\text{initial}}$ to $\sigma_{\text{final}}$ over training epochs, enabling coarse-to-fine OT alignment.

  \item \textbf{Domain-adversarial training (DANN).} As a complementary baseline, we implement a GRL-based domain discriminator \cite{dann} acting on $(z_s,z_t)$ (\texttt{DAAdversarialTrainer}).  
  The GRL scale $\lambda_{\text{adv}}$ is tied to \texttt{lambda\_da} (set to $0.1$ in the main experiments).
  
\end{itemize}


\paragraph{Top-$k$ soft domain alignment.} Using the composite OT loss defined above, we test a restricted top-$k$ / soft-matching variant that focuses the alignment on the hardest-to-match examples.
Standard batch-level alignment treats all source–target pairs equally, which can encourage mismatched couplings (e.g. spirals aligned to ellipticals) when the domains differ.
Our OT term instead combines a transport-induced soft matching with a sparse top-$k$ penalty on the worst-aligned source features, encouraging ellipticals to align with ellipticals and spirals with spirals rather than with arbitrary batch neighbors.
This is the key modification that yields our best domain-alignment performance in our experiments, including all three backbones.

\paragraph{Training procedure.}
All models are trained for up to 200 epochs with early stopping based on the training loss (patience of 10 epochs). We use a batch size of 128 and resize all images to $224{\times}224$ pixels. For domain-adaptation methods, we include a warmup phase of 20 epochs during which only the supervised focal loss is applied, allowing the model to establish a reasonable feature representation before introducing domain-alignment terms. We optimize using AdamW with a weight decay of $10^{-2}$ and apply gradient clipping with a maximum norm of 10.0 to stabilize training. The learning rate follows a cosine-annealing schedule, starting at $10^{-4}$ and decaying to $10^{-6}$ over the full training horizon, reducing the step size as training progresses while preserving relatively large updates early on. Source and target images are normalized separately using domain-specific statistics computed from their respective training sets, preserving the natural distributional differences that domain adaptation aims to bridge. 

\section{Results}

\begin{figure}[t!]
    \centering
    \begin{minipage}[b]{0.58\textwidth}
        \centering
        \includegraphics[width=\linewidth, height=5cm, keepaspectratio]{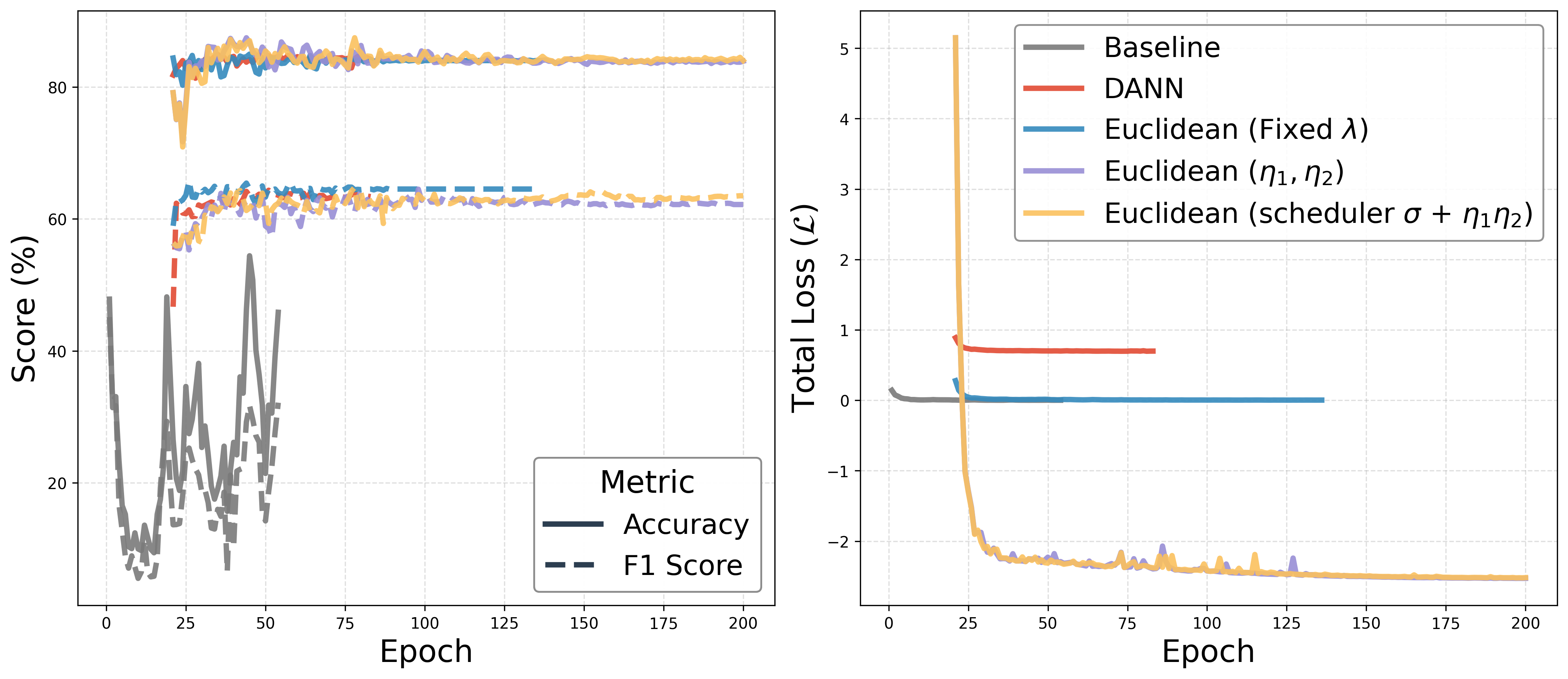}
        \captionof{figure}{\textbf{Training Dynamics.} Comparison of stability and loss convergence.}
        \label{fig:perf_epoch}
    \end{minipage}
    \hfill 
    \begin{minipage}[b]{0.38\textwidth}
        \centering
        \includegraphics[width=\linewidth, height=5cm, keepaspectratio]{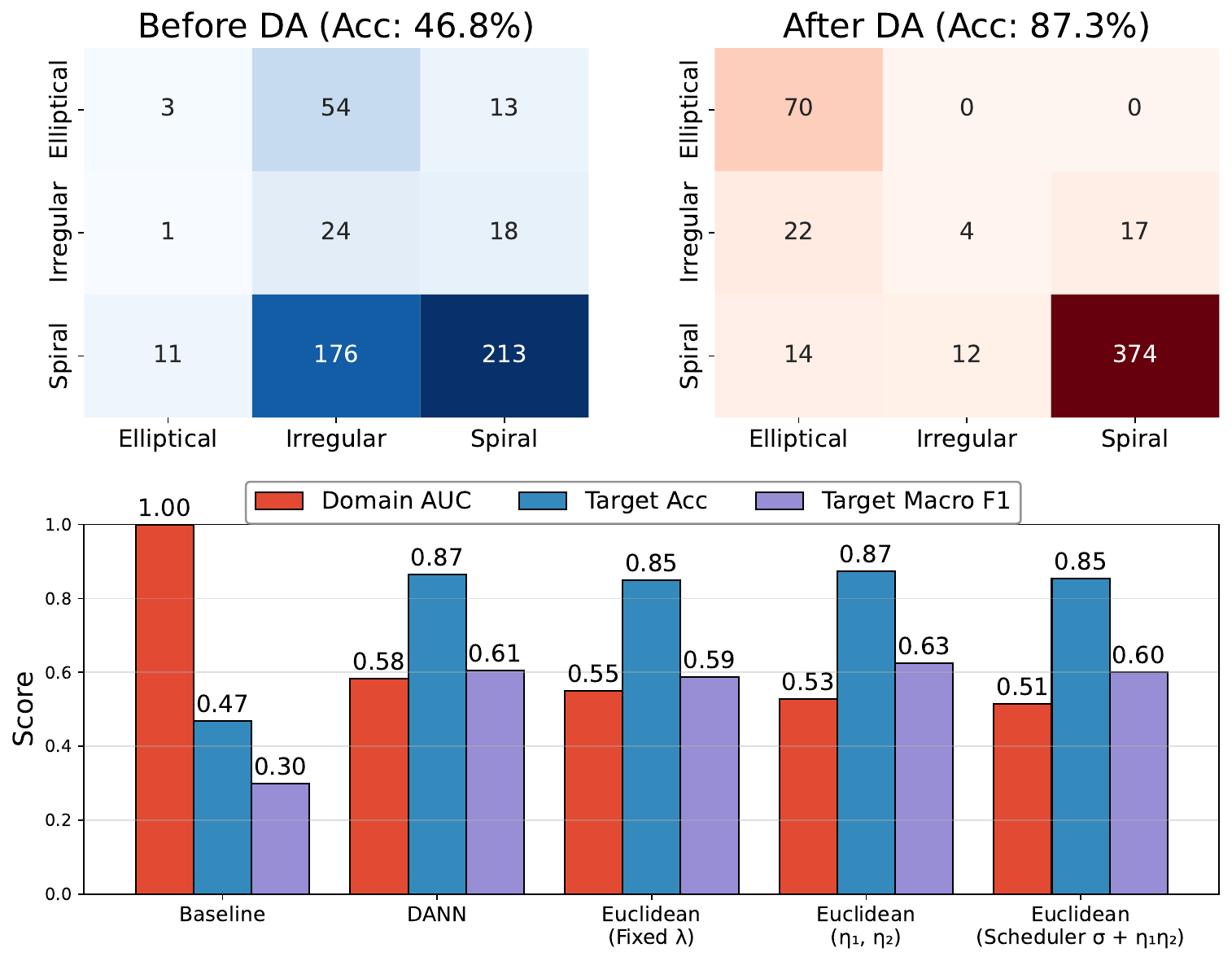}
        \captionof{figure}{\textbf{Target Accuracy.} Confusion matrix and metrics.}
        \label{fig:confusion_matrix}
    \end{minipage}
    
    \vspace{2mm} 

    \includegraphics[width=0.7\textwidth]{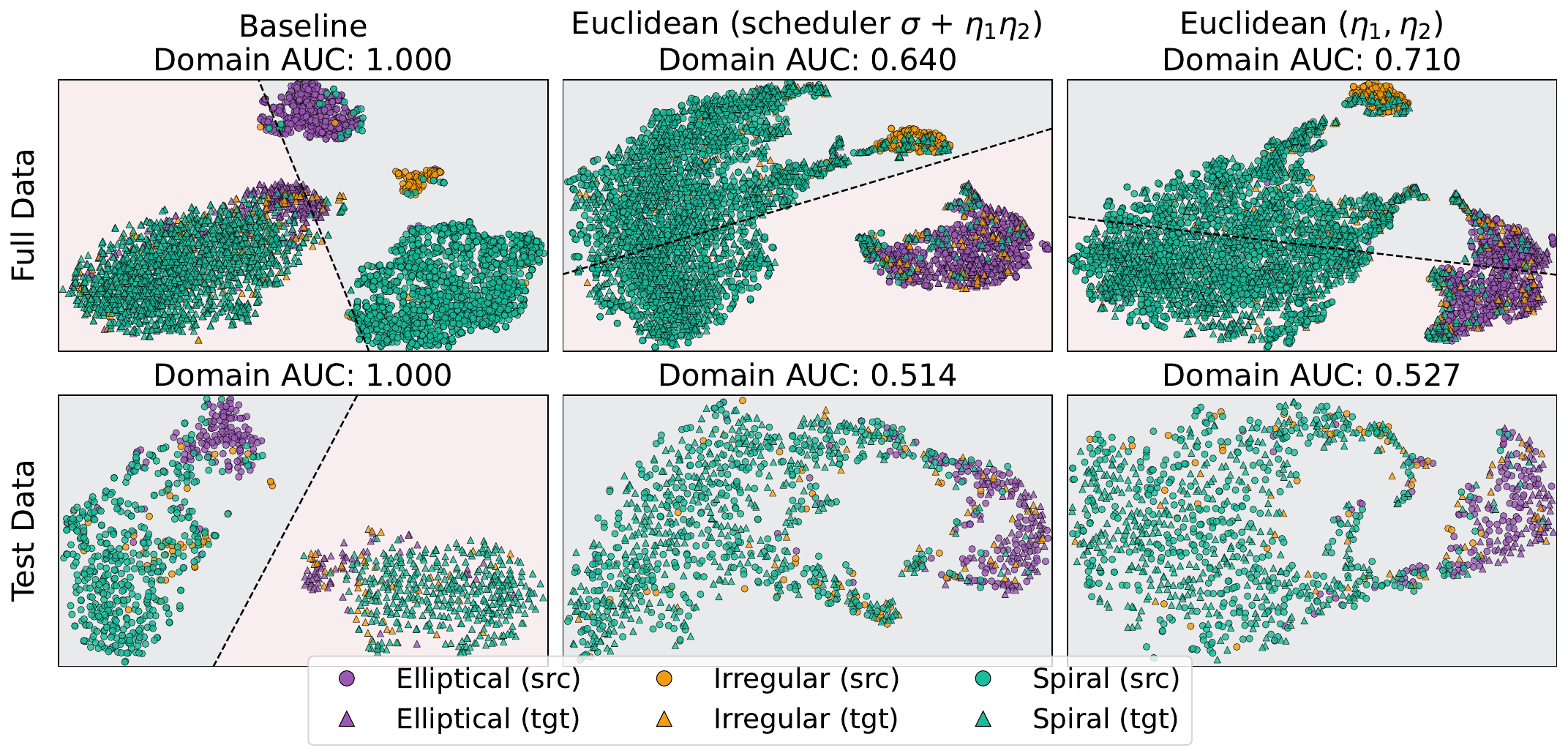}
    \vspace{-1mm}
    \caption{\textbf{Latent space embedding.} The Baseline (left, AUC=1.00) shows distinct domain separation, while Euclidean variants (center/right, AUC$\approx$0.51) achieve effective domain alignment where source and target distributions are indistinguishable. In an ideal domain-invariant representation, AUC = 0.5.}
    \label{fig:latent_space}
    \vspace{-4mm} 
\end{figure}

\paragraph{Accuracy improvements on target domain.}
Figure~\ref{fig:perf_epoch} illustrates 200-epoch training dynamics for baseline, adversarial, and Euclidean models, selecting Euclidean as a representative metric for brevity. The left panel displays target-domain accuracy (solid lines) and macro–F1 (dashed lines). The performance trajectories reveal a clear distinction: the Baseline (grey) exhibits significant instability and terminates early due to loss degradation, whereas the domain adaptation methods show a sharp performance jump after the 20-epoch warmup phase, maintaining high stability thereafter. The loss plot (right panel) further distinguishes the trainable Euclidean variants (purple and yellow), which converge to a distinct negative loss range compared to the fixed parameters.

At peak performance (epoch 197), the Euclidean method with trainable weights ($\eta_1, \eta_2$) achieved the best results (\textbf{87.3\% accuracy, 0.626 macro–F1}), a dramatic improvement over the baseline (46.8\%, 0.298). This quantitative boost is visualized in the confusion matrices (Fig.~\ref{fig:confusion_matrix}), which demonstrate significantly cleaner class separation after adaptation. 
While DANN (86.5\%) and other Euclidean variants yielded competitive gains, the fully trainable weighting scheme consistently provided the most stable alignment. We find that top-$k$ soft-matching with ResNet effectively aligns classes across domains, driving these improvements in both accuracy and F1.

\section{Discussion \& Next Steps}

This work is the precursor to a domain-adapted predictive model to leverage hundreds of thousands of mock observations from the Illustris simulations at different redshifts with the ultimate goal of interpreting upcoming Rubin galaxy observations.

Next steps include 
(1) \textbf{Richer labels:} extending from three-way morphology to a multi-task setting (e.g., star-forming vs. quenched, AGNs, stellar mass.) 
(2) \textbf{Irregular class:} improving the rare irregular class with targeted augmentations and stronger imbalance handling. 
(3)  \textbf{Latent space and distance metrics:} probing the learned feature space under different distance metrics in $\mathcal{L}_D$ to understand when particular metrics yield better physical alignment between simulations and observations, and to guide the choice of
future architectures.
(4) \textbf{Learning rate scheduler:} incorporating distance-aware schedulers that modulate the learning rate using alignment signals such as Sinkhorn cost, top-$k$ distances, or barycentric mismatch, and quantifying their impact on convergence stability and domain-alignment.
(5) \textbf{Alternative architectures:} testing equivariant transformers or graph-based models to see how different inductive biases affect the geometry of the learned latent space.

\begin{ack}
This publication uses data generated via the Zooniverse.org platform, development of which is funded by a Global Impact Award from Google, and by a grant from the Alfred P. Sloan Foundation. K.B. is supported by an NSF Astronomy and Astrophysics Postdoctoral Fellowship under award AST-2303858. A.P.D. is supported by Teaching Fellowship and Physics Division Fellowship at UCLA and US DOE Office of Science Grant No. DE-FG02-88ER4. We are grateful to the IAIFI Summer School organizers, where this project first took shape.
\end{ack}


\bibliography{refs}





\end{document}